\documentclass[english,keywords,amsmath,amssymb,twocolumn]{revtex4}
\usepackage[T1]{fontenc}
\usepackage[latin1]{inputenc}
\usepackage{braket}
\usepackage{babel}%
\usepackage{graphicx}
\usepackage{color}
\usepackage{bm}
\usepackage{longtable}
\usepackage{amsmath}
\usepackage{amsfonts}
\usepackage{dsfont}
\usepackage{amssymb}
\usepackage{hyperref}
\begin{document}
\title{Sharing non-locality and non-trivial preparation contextuality using same family of Bell expressions }
\author{Asmita Kumari}
\author{A. K. Pan \footnote{akp@nitp.ac.in}}
\affiliation{National Institute of Technology Patna, Ashok Rajhpath, Patna 800005, India}
\begin{abstract}
 In [Phys. Rev. Lett. 114, 250401 (2015)] the sharing of non-locality by multiple observers was demonstrated through the quantum violation of Clauser-Horne-Shimony-Halt inequality. In this paper we provide a scheme for sharing of non-locality and non-trivial preparation contextuality sequentially through the quantum violation of a family of Bell's inequalities where Alice and Bob perform  $2^{n-1}$ and $n$ numbers of  measurements of dichotomic observables respectively.  For this, we consider that Alice always performs projective measurement and multiple Bobs sequentially perform unsharp measurement. We show that when Bob's choices of measurement settings are unbiased, maximum two Bobs can  sequentially share the non-locality through the violation of our inequalities. Further, we show that the local bound of the aforementioned family of inequalities gets reduced if non-trivial preparation non-contextuality assumptions are further imposed. Then there is a chance to share the non-trivial preparation contextuality for more number of Bobs than that of non-locality. We demonstrate that the non-trivial preparation contextuality can be sequentially shared by arbitrary numbers of Bob for unbiased choices of his measurement settings.  
 
\end{abstract}
\maketitle
\section{Introduction}
No-go theorems play a pivotal role in quantum foundational research. They provide an elegant route to discriminate the various notions of classical physics from quantum theory. Bell's no-go theorem \cite{bell64} is arguably the most famous one which first shows that all statistics of quantum theory cannot be reproduced by any theory that respects a notion of classicality widely known as local realism. Another no-go theorem was provided by Kochen and Specker (KS) \cite{ks} to demonstrate the incompatibility of quantum theory with non-contextual realism.  One of the most fundamental distinctions between the quantum and classical theory is the concept of measurement. Unlike that in classical theory, the measurement in general disturbs the pre-measurement state of the quantum system.  And a projective measurement disturbs  quantum system most by collapsing the initial state of the system into one of the eigenstates of the measured observable. In other words, in a von Neumann type ideal (sharp) projective measurement scenario \cite{von,home}, the information gain about the system is maximum but the memory of pre-measurement state is completely lost, thereby occurring maximum disturbance to the system. However, in an unsharp measurement scenario, where system is not fully disturbed, a partial coherence remains in the system. In such a case, the residual coherence can be revealed by subsequent measurement. 

In a bipartite Bell scenario where Alice and Bob share a suitable entangled state, the non-local correlation can be revealed through the quantum violation of suitable Bell's inequality. The simplest Bell's inequality in bipartite binary outcome scenario is the Clauser-Horne-Shimony-Holt (CHSH) form \cite{chsh69}. Recently, an important question was put forwarded by Silva \emph{et al.}\cite{silva} that if an entangled pair of particles is shared between a single Alice and multiple Bobs and each Bob performs unsharp measurement on the same particle, then how many Bob can sequentially share the non-locality by demonstrating the violation of CHSH inequality. It is straightforward to understand that if first Bob's measurement is projective, the entangled state becomes an unentangled one, and hence there is no way to get violation of a Bell's inequality  for later Bobs. But, if first Bob performs the unsharp measurement on his particle by suitably choosing POVMs, then due to the presence of residual entanglement there is a chance to share the non-locality to the next Bob through the violation of a suitable Bell's inequality. Higher the disturbance caused by the first Bob, lower the correlation remains for second Bob and so on. Thus, sharing the non-locality by larger number of sequential observers requires that the Bob's measurements to be as unsharp  as possible but enough for violating the relevant Bell's inequality. 

For two-qubit entangled sysem, Silva \emph{et al.}\cite{silva} demonstrated that non-locality through the violation CHSH inequality can be shared by at most two Bobs for the unbiased choices of measurement settings. However, if the choices of measurement settings are taken to be biased from second Bob, then the sharing of non-locality can be demonstrated for arbitrary sequences of Bob. Their result has recently been verified experimentally \cite{schiavon,hu}. This idea is further extended \cite{malar,das} for other form of Bell inequalities where Alice and Bob measure more than two observables (instead of two observables measured by each Alice and Bob in CHSH scenario) but  sharing of non-locality is restricted to two Bobs for unbiased settings of Bob. Further, the study of the sharing of entanglement \cite{bera}, steering \cite{sasmal}, coherence \cite{datta} has also been reported. It has also recently been demonstrated that unbounded number of Bobs can steer Alice's  state \cite{shenoy} for a suitable choice of entangled state in higher dimension.  

In the present work, we examine the sharing of non-locality and non-trivial preparation contextuality (introduced in detail in Sec. II) through the quantum violation of a family of Bell's inequalities \cite{ghorai18} where Alice and Bob use $2^{n-1}$ and $n$ number of dichotomic measurements respectively. We consider that the measurement settings chosen by each Bob (from second Bob) are unbiased. In a bipartite binary measurement Bell scenario, any proof of non-locality is in fact a proof of \emph{trivial} preparation contextuality \cite{pusey}. Given the family of  Bell's local realist inequalities, if the assumption of non-trivial preparation non-contextiuality  is additionally imposed, the local realist inequalities becomes  non-trivial preparation non-contextual  inequalities. Such assumption of non-trivial preparation non-contextiuality originates from the functional relations between the observables belong to one of the two parties (say, Alice). In other words, the local bound of the aforementioned family of Bell expressions  get reduced to non-trivial non-contextual bound if the a constraint on Alice's observables are further imposed. However, for CHSH scenario no such non-trivial relations between Alice's observables can be found and so that, the local and non-trivial preparation non-contextual bound of CHSH expression remain same.   

Since the non-trivial preparation non-contextual bound of the family of inequalities is less than the local bound (as explicitly shown in Sec. II), then for a given entangled state and choices of measurements, the quantum violation of the former may not always imply the violation of  later. That is, the quantum violation of local bound always imply the demonstration of both trivial as well as non-trivial preparation contextuality but converse may not be true. This provides an indication that non-trivial preparation contextuality can be shared for a larger number of Bobs than that of non-locality. We specifically demonstrate that for the family of Bell expressions considered here, even though the non-locality can be shared only by two sequential Bobs, the non-trivial preparation contextuality can be shared by any arbitrary number of Bobs even when his choices of measurement settings are unbiased.  However, in such a case the dimension of the Hilbert space is required to be $d= 2^{\lfloor n/2\rfloor} $. We study this issue by considering that Bob performs one- and two-parameter family of POVMs. It is found that for both forms of POVMs, the sharing of non-trivial preparation contextuality can be demonstrated for any arbitrary number of Bobs (k) for which $n$ has to be grater than or equal to $k$.  

 This paper is organized as follows. In Sec.II, we demonstrate the local and non-trivial non-contextual bound of a family of Bell's expressions. In Sec.III, we calculate the optimal quantum value of family of Bell's inequality for specific choice of two-parameter POVMs while Bob performs any arbitrary number of sequential unsharp measurements.  Further, we derive the condition on unsharpness parameter for sharing of non-locality and preparation contextuality among multiple observers  in Sec. IV and Sec. V respectively, by considering both one- and two-parameter POVMs. It is found that Alice can share the preparation contextuality with any arbitrary number of Bobs. A discussion regarding the biased choice of settings by Bob is provided in Sec. VI.  We discuss our results in Sec. VII. 

\section{A family of Bell's inequalities and their local and preparation non-contextual bound} 

In order to derive the local and preparation non-contextual bound of the aforementioned family of Bell expression, we start by encapsulating the notion of an ontological model reproducing the quantum statistics \cite{hari}. Given a preparation procedure $P$ and a measurement procedures $M$, an operational theory assigns probability $p(k|P, M)$ of obtaining a particular outcome $k$. Here $\mathcal{M}$ is the set of measurement procedures and $\mathcal{P}$ is the set of preparation procedures. In quantum mechanics (QM), a preparation procedure produces a density matrix $\rho$ and measurement procedure (in general described by a suitable POVM $E_k$) provides the probability of a particular outcome $ k $ is given by $p(k|P, M)=Tr[\rho E_{k}]$, which is the Born rule. In an ontological model of QM, it is assumed that whenever $\rho$ is prepared by a specific preparation procedure $P$ a probability distribution $\mu_{P}(\lambda|\rho)$ in the ontic space is prepared, satisfying $\int _\Lambda \mu_{P}(\lambda|\rho)d\lambda=1$ where $\lambda \in \Lambda$ and $\Lambda$ is the ontic state space. The probability of obtaining an outcome $k$ is given by a response function $\xi_{M}(k|\lambda, E_{k}) $ satisfying $\sum_{k}\xi_{M}(k|\lambda, E_{k})=1$ where a measurement operator $E_{k}$ is realized through a particular measurement procedure $M$. A viable ontological model should reproduce the Born rule, i.e., $\forall \rho $, $\forall E_{k}$ and $\forall k$, $\int _\Lambda \mu_{P}(\lambda|\rho) \xi_{M}(k|\lambda, E_{k}) d\lambda =Tr[\rho E_{k}]$.

The notion of non-contextuality was reformulated and generalized by Spekkens\cite{spekk05}.  An ontological model of an operational theory can be assumed to be non-contextual if two experimental procedures are operationally equivalent then they have equivalent representations in the ontological model.  If two measurement procedures $ M $ and $ M' $ produces same observable statistics for all possible preparations then the measurements $M$ and $M'$ belong to the equivalent class. An ontological model of QM is assumed to be measurement non-contextual if $\forall P :  p(k|P, M)=p(k|P, M^{\prime})\Rightarrow \xi_{M}(k|\lambda, E_{k})=\xi_{M^{\prime}}(k|\lambda, E_{k})$ is satisfied. KS non-contextuality assumes the aforementioned measurement non-contextuality along with the outcome determinism for the sharp measurement. The traditional notion of KS non-contextuality was generalized by Spekkens\cite{spekk05} for any arbitrary operational theory and extended the formulation to the transformation and preparation non-contextuality.  An ontological model of QM can be considered to be preparation non-contextual if  $\forall M :  p(k|P, M)=p(k|P^{\prime}, M)\Rightarrow \mu_{P}(\lambda|\rho)=\mu_{P^{\prime}}(\lambda|\rho)$ is satisfied where $P$ and $P^{\prime}$ are two distinct preparation procedures but in the same equivalent class.

We derive a suitable family of Bell's expressions by using a multiplexing game as a tool. If parity-oblivious conditions are further imposed in that communication game, then such conditions will have equivalent representation in the ontological model and are regarded as the preparation non-contextual assumptions. The essence of a $n$-bit  parity-oblivious multiplexing (POM) game can be encapsulated as follows. Alice has a $ n $-bit string $ x^{\delta} $ with $\delta  \in \{1,2, ...2^{n}\}$ chosen uniformly at random from $\{0,1\}^n$. The relevant ordered set $\mathcal{D}_n$ can be written as 
$\mathcal{D}_n = (x^{\delta} | x^i \oplus x^l = 111...11 \;\text{and}\; i+l = 2^n +1 ) $
and $i \in \{1,2, ...2^{n-1}\}$. Here, $ x^1 = 00...00, x^2 = 00...01, .... $, and so on. Bob can choose any bit $ y \in \{1,2, ..., n\}$ and recover the bit $x^{\delta}_y$ with a probability. The condition of the task is, Bob's output must be the bit $b=x^{\delta}_y$, i.e., the $ y^{th} $ bit of Alice's input string $x$ with the constraint that \emph{no} information about any parity of $ x $ can be transmitted to Bob. Following \cite{spekk09}, we define a parity set $ \mathbb{P}_n= \{x|x \in \{0,1\}^n,\sum_{r} x_{r} \geq 2\} $ with $r\in \{1,2,...,n\}$. For any arbitrary $s \in \mathbb{P}_{n}$, no information about $s.x = \oplus_{r} s_{r}x_{r}$ (s-parity) is to be transmitted to Bob, where $\oplus$ is sum modulo $ 2 $. 

In QM, Alice encodes her $ n $-bit string of $ x^{\delta} $ into a density matrix $\rho_{x^{\delta}}$ prepared by a procedure $P_{x^{\delta}}$.  Let us consider a suitable entangled state $ \rho_{AB} = \ket{\phi_{AB}}\bra{\phi_{AB}}$ with $|\phi_{AB}\rangle \in \mathcal{C}^{d}\otimes\mathcal{C}^{d}$. Alice performs one of the $2^{n-1}$ projective measurements $\{{P_{A_{n,i}}}, \mathbb{I}-P_{A_{n,i}}\}$  to encode her $n$-bits into $ 2^n $ density matrices are given by
\begin{subequations}
	\begin{align}
	\dfrac{1}{2} \rho_{x^i} &= tr_A[(P_{A_{n,i}} \otimes \mathbb{I}) \rho_{AB}]\\
	\dfrac{1}{2} \rho_{x^l} &= tr_A[(\mathbb{I}-P_{A_{n,i}} \otimes \mathbb{I}) \rho_{AB}] 
	\end{align}
\end{subequations}
with $ i+l=2^n+1 $.

The parity-obliviousness condition in QM can mathematically be written as
\begin{equation}
\label{poc}
\forall s \ \ \ \  \frac{1}{2^{n-1}}\sum\limits_{x^{\delta}|x^{\delta}.s=0} \rho_{x^{\delta}}=\frac{1}{2^{n-1}}\sum\limits_{x^{\delta}|x^{\delta}.s=1} \rho_{x^{\delta}}
\end{equation} 
In ontological model of QM, parity-obliviousness in Eq.(\ref{poc})  condition provides preparation non-contextual assumption, i.e., 

\begin{align}
	\forall s: \frac{1}{2^{n-1}}\sum\limits_{x^{\delta}|x^{\delta}.s=0} \mu(\lambda|\rho_{x^{\delta}})=\frac{1}{2^{n-1}}\sum\limits_{x^{\delta}|x^{\delta}.s=1} \mu(\lambda|\rho_{x^{\delta}})
\end{align}

Note that the number of parity-oblivious conditions for $n$-bit POM task is the number of element in $ \mathbb{P}_n$ \cite{ghorai18}. We noticed that there are two types of parity oblivious conditions. The one arising from the natural construction, such as, $\mathbb{I}=P_{A_{n,i}}^{+}+P_{A_{n,i}}^{-}$. In that case, $s \in \mathbb{P}_n $ follow the property $ \sum_y s_y = 2m$ with $m \in \mathbb{N} $.  For the rest of $s \in \mathbb{P}_n$ not satisfying the above property, a non-trivial constraints on Alice's observables need to be satisfied are given by

\begin{equation}
\label{oba}
\sum_{i=1}^{2^{n-1}} (-1)^{s.x^i} A_{n,i} = 0
\end{equation}
The total number of such non-trivial constraints on Alice's observables is  $C_n= 2^{n-1} - n$.

Since parity-oblivious conditions in QM is equivalent preparation non-contextuality assumptions in ontological model, we then have trivial and non-trivial preparation non-contextuality assumptions. 

Now, for every $y \in \{1,2,...,n\}$, Bob performs a two-outcome measurement $M_{n,y}$ and reports outcome $b$ as his output. Bob's measurements are the following.  
\begin{align}
M_{n,y} = \begin{cases}
M_{n,y}^i,\text{when}\; b=x_y^i \\ M_{n,y}^l, \text{when}\; b=x_y^l
\end{cases}\\
M_{n,y}^{i(l)} = \begin{cases}
P_{B_{n,y}},& \text{when}\; x^{i(l)}_y=0 \ \\ \mathbb{I} - P_{B_{n,y}},& \text{when}\; x^{i(l)}_y=1 
\end{cases}	
\end{align}

The average success probability in QM can then be written as 
\begin{align}
\nonumber
p_Q & =\dfrac{1}{2^n n} \sum_{y=1}^{n} \sum\limits_{i=1}^{2^{n-1}} tr[\rho_{x^i} M_{n,y}^i] + tr[\rho_{x^l} M_{n,y}^l] \\ 
& = \frac{1}{2} + \frac{1}{2^n n} \langle \mathcal{B}_{n} \rangle_{Q}
\end{align} 
where $\langle\mathcal{B}_{n} \rangle_{Q}$ is a family of Bell expressions, so that  
\begin{align}
\label{nbell1}
\mathcal{B}_{n} =  \sum_{y=1}^{n}\sum_{i=1}^{2^{n-1}} (-1)^{x^i_y}  A_{n,i}\otimes B_{n,y}
\end{align}
For $n=2$ and $3$, the Bell expression $\mathcal{B}_{n}$ become the well-known CHSH \cite{chsh69} and Gisin's elegant Bell \cite{gisin} expressions. Using sum of square decomposition \cite{ghorai18} the maximum quantum value we have $\langle\mathcal{B}_{n} \rangle_Q \leq 2^{n-1}\sqrt{n}$. In order to obtain optimal quantum value, the choices of observables has to satisfy $(2^{n-1}/\sqrt{n})\sum_{i=1}^{2^{n-1}} (-1)^{x^i_y}  A_{n,i} \otimes \mathbb{I}= B_{n,y}\otimes \mathbb{I}$ \cite{ghorai18} and requires a maximally entangled state of the dimension  $ 2^{\lfloor n/2\rfloor} $ is of the form given by
\begin{equation}
\label{ent}
|{\phi}\rangle_{AB} = \dfrac{1}{\sqrt{2^{\lfloor n/2\rfloor}}} \sum\limits_{k=1}^{2^{\lfloor n/2\rfloor}}  |{k}\rangle_A |{k}\rangle_B .
\end{equation}
For $n-$ bit case, Bob requires $n$ number of mutually unbiased basis. For 2-bit case $B_{2,1} = \sigma_x, B_{2,2} = \sigma_y $, for 3-bit case $ B_{3,1} = \sigma_x, B_{3,2}=\sigma_y ,B_{3,3} = \sigma_z $, and for 4-bit case one requires $ B_{4,1} = \sigma_x \otimes \sigma_x , B_{4,2} = \sigma_x \otimes \sigma_y ,B_{4,3} = \sigma_x\otimes \sigma_z ,B_{4,4} =  \sigma_y\otimes\mathbb{I}$. Using $n=3$ and $4$ cases, one finds the required observables. For even  $n$, $B_{n,y} = \sigma_x \otimes B_{n-1,y} \;\text{for} \; y \in \{1, . . ., n-1\}, \quad B_{n,n} = \sigma_y \otimes I $ and for odd $n$, $B_{n,y} = \sigma_x \otimes B_{n-2,y} \;\text{for} \; y \in \{1, . . ., n-2\}, \quad B_{n,n-1} = \sigma_y \otimes I \; \text{and} \;  B_{n,n} = \sigma_z \otimes I $.

In order to demonstrate the sharing of non-locality and the non-trivial preparation contextuality, we use the family of Bell's expression derived in Eq.(\ref{nbell1}).  The local bound of it is derived as
\begin{align} 
\label{localbound}
	{(\mathcal{B}_{n})}_{local} \leq n\binom{n-1}{\lfloor\frac{n-1}{2}\rfloor}  
\end{align}
For $n=2$ and $3$ we have,  ${(\mathcal{B}_{n})}_{local}\leq 2$ and $6$ respectively.

Now, in order to satisfy the parity-obliviousness condition in QM for $n$-bit case, total $C_n= 2^{n-1} - n$ number of non-trivial relations between the Alice's observables given by Eq.(\ref{oba}) need to be satisfied \cite{ghorai18}. For example, when $n=3$ there is only one constraint $A_{4,1}- A_{4,2}- A_{4,3}- A_{4,4}=0$, where $A_{4,i}$ $(i=1,2,3,4)$ are the Alice's observables. Interestingly, for $n=2$, when ${\mathcal{B}_{2}}$ in Eq. (\ref{nbell1})  is just the CHSH expression, no such non-trivial constraint can be found. As already stated earlier that  for two-party, two measurement per site and two-outcome Bell scenario the trivial preparation non-contextuality is in fact the assumption of locality \cite{barrett,pusey}. Thus, local and preparation non-contextual bound are the same for CHSH expression. 

 As already mentioned, the parity-oblivious conditions in Eq.(\ref{oba}) will have equivalent representation in an ontological model, which in turn provide the trivial and non-trivial preparation non-contextuality assumptions. Using the non-trivial relations given by Eq.(\ref{oba}), from Eq.(\ref{nbell1}) we have
\begin{align}
\label{npnc}
	(\mathcal{B}_{n})_{pnc} \leq  2^{n-1}
\end{align}
which is family of non-trivial preparation non-contextual inequalities.

Since $(\mathcal{B}_{n})_{local}> 	(\mathcal{B}_{n})_{pnc}$ for any $n>2$, then  even if the quantum value $(\mathcal{B}_{n})_{Q}$ is not enough to exhibit the violation of local bound in Eq.(\ref{localbound}) there is still a chance of violating the non-contextual bound in Eq.(\ref{npnc}). In other words, when $ (\mathcal{B}_{n})_{local} \geq (\mathcal{B}_{n})_{Q}	> (\mathcal{B}_{n})_{pnc} $, although no non-local correlation is revealed but a non-classicality in the form of non-trivial preparation contextuality can be revealed. One may then notice that there is a chance of sharing that non-trivial preparation contextual correlation for more number of Bobs than non-locality. We show that any arbitrary number of Bobs can sequentially share the non-trivial preparation contextuality by violating Eq.(\ref{npnc}), depending upon the dimension of the Hilbert space. But sharing non-locality through the family of Bell's inequality given by Eq.(\ref{npnc}) is restricted to two Bobs only.

\section{Quantum value of Bell expression for arbitrary sequential Bob}
 To find the number of sequential Bobs sharing non-locality and the non-trivial preparation contextuality, let us consider that there is only one Alice who performs sharp measurement and $k$ number of Bobs perform unsharp measurement sequentially. However, the $k^{th}$ Bob may perform a projective measurement.  For the Bell expression $\mathcal{B}_{n}$ Alice and each Bob  perform the measurements $2^{n-1}$ and $n$ number of dichotomic observable respectively. Given a $n$ value, each Bob requires to perform same set of $n$ number of observables for every $j$. We also consider that Bob's choices of measurement settings are completely random. For example in $n=2$ case, for any of the  two observables (say, $B_{2,1}$) chosen by first Bob, the second Bob randomly performs $B_{2,1}$ and $B_{2,2}$ on the reduced state which is  obtained from the first Bob's unsharp measurement of $B_{2,1}$.  

Note that the sharing of non-locality and non-trivial preparation contextuality for larger number of Bobs is expected when the quantum value of $(\mathcal{B}_{n})_{Q}$ is maximum. We have already shown in Sec. II that there are specific choices of observables of Alice and Bob for which $(\langle\mathcal{B}_{n} \rangle_Q)_{max} = 2^{n-1}\sqrt{n}$ . We use the same set of observables for Alice and $k^{th}$ Bob for examining the sharing of the non-locality and non-trivial preparation contexuality. 
 
  Let Alice and Bob$_1$ share a maximally entangled state given by Eq.(\ref{ent}) and each of the $k-1$ number of Bobs performs the measurement of two-parameter dichotomic POVMs is given by
\begin{align}
\label{maxe}
{E}^{\pm}_{B_{n,y,j}}=    \frac{1 \pm \alpha_{n,j} \pm \eta_{n,j}}{2} \Pi^{+}_{B_{n,y}}  + \frac{1 \pm \alpha_{n,j} \mp \eta_{n,j}}{2} \Pi^{-}_{B_{n,y}} 
\end{align}
 where $ \Pi^{\pm}_{B_{n,y}}$ are the projectors of Bob's observable $B_{n,y}$ with $y = (1,2,....,n)$. Here $\eta_{n,j}$ $(0<\eta_{n,j}\leq1)$ and $\alpha_{n,j}$ $(0\leq|\alpha_{n,j}|\leq1)$ are the sharpness and biasedness parameters satisfying $|\alpha_{n,j}|+\eta_{n,j} \leq 1$ of $j^{th}$ Bob where $j = 1,2....k$ \cite{busch, kumari19}. Here we consider that the unsharpness and biasedness parameters are same for all observables of Bob for a given $n$ and thus independent of $y$ in Eq.(\ref{maxe}). For example, for $n=2$, Bob performs the measurement of two observables $B_{2,1}$ and  $B_{2,2}$. We consider the same unsharpness parameter $\eta_{2,1}$ for the unsharp measurements of both $B_{2,1}$ and  $B_{2,2}$ of first Bob, $\eta_{2,2}$ for the second Bob and so on. Similarly for the biasedness parameter.

The shared state between Alice and ${k}^{th}$ Bob is obtained after the unsharp  measurements of $k-1$ Bobs is given by
\begin{eqnarray}
\nonumber
\rho_{n,k}  &=& \frac{1}{n}\sum_{b \in \left\{ + ,- \right\}}\sum_{y = 1}^{n} (\mathbb{I} \otimes \sqrt{ {E}^{b}_{B_{n,y,k-1}}}) \rho_{n,k-1} (\mathbb{I} \otimes \sqrt{ {E}^{b}_{B_{n,y,k-1}}})  \\ \nonumber 
&=& \xi_{n,k-1}\rho_{n,k-1}+\frac{(1- \xi_{n,k-1})}{n}\times\\
&&\sum_{b \in \left\{ + ,- \right\}}\sum_{y = 1}^{n} (\mathbb{I} \otimes {\Pi}^{b}_{B_{n,y,k-1}}) \rho_{n,k-1} (\mathbb{I} \otimes  {\Pi}^{b}_{B_{n,y,k-1}})
\end{eqnarray}
where $\rho_{n,k-1}$ is the state shared between Alice and ${(k-1)}^{th}$ Bob before $(k-1)^{th}$ Bob's unsharp measurement and
\begin{eqnarray}
\xi_{n,k-1} &=& \frac{1}{2}\bigg[\sqrt{(1+\alpha_{n,k-1})^2-\eta_{n,k-1}^2}\\ \nonumber
	&+&\sqrt{(1-\alpha_{n,k-1})^2-\eta^2_{n,k-1}}\bigg]
\end{eqnarray}

For a maximally entangled state  given by Eq.(\ref{ent}), the Bell expression in Eq.(\ref{nbell1}) for $k^{th}$ Bob is obtained as
\begin{eqnarray}
\label{quant}
\nonumber
(\mathcal{B}^{k}_n)_Q &=& 2^{n-1}\sqrt{n} \bigg[\bigg(1+(n-1)\xi_{n,1}\bigg) \bigg(1+(n-1)\xi_{n,2} \bigg)....\\ \nonumber &&....\bigg(1+(n-1)\xi_{n,k-1}\bigg)\bigg]\eta_{n,k}\\  &=& 2^{n-1}\sqrt{n}\eta_{n,k} \prod^{k-1}_{j=1} \gamma_{n,j} 
\end{eqnarray}
where 
\begin{eqnarray}
\label{ga}
\gamma_{n,j}=(1+(n-1)\xi_{n,j})/n
\end{eqnarray}
with $\xi_{n,j}=(\sqrt{(1+\alpha_{n,j})^2-\eta_{n,j}^2}+\sqrt{(1-\alpha_{n,j})^2-\eta^2_{n,j}})/2$. 

Note here that, for first Bob $(k=1)$, the value of $(\mathcal{B}^{1}_n)_Q=2^{n-1}\sqrt{n}\ \eta_{n,1}$. For  $n=2$ case, we have the known results of the optimal quantum value of CHSH expression $2\sqrt{2} \ \eta_{2,1}$.

 The non-locality and non-trivial preparation contextuality can be shared by $k^{th}$ Bob, if $(\mathcal{B}^{k}_n)_Q>(\mathcal{B}_n)_{local}$ and  $(\mathcal{B}^{k}_n)_Q>(\mathcal{B}_n)_{pnc}$ are respectively satisfied. Our purpose will be to find the legitimate value of sharpness and biasedness parameters for $k^{th}$ Bob while satisfying the condition $(\mathcal{B}^{k}_n)_Q>(\mathcal{B}_n)_{local}$ for sharing non-locality and  the condition $(\mathcal{B}^{k}_n)_Q>(\mathcal{B}_n)_{pnc}$ for sharing non-trivial preparation contextuality.

\section{Sharing non-locality for $k^{th}$ sequential Bob}

We first examine how many Bobs can sequentially share non-locality through the violation of Eq.(\ref{localbound}). The sharing of non-locality for $k^{th}$ Bob requires the condition $(\mathcal{B}^{k}_n)_{Q} > n\binom{n-1}{\lfloor\frac{n-1}{2}\rfloor}  $ to be satisfied. From Eq.(\ref{quant}) one can then write the general condition on sharpness parameter $\eta_{n,k} $ for $n$-bit case is given by
 \begin{eqnarray}
\label{lvcond}
\eta_{n,k}  > \frac{n\binom{n-1}{\lfloor\frac{n-1}{2}\rfloor}}{2^{n-1} \sqrt{n}\prod^{k-1}_{j=1} \gamma_{n,j}}
\end{eqnarray} 
We examine upto what value of $k$ the condition in Eq.(\ref{lvcond}) is satisfied by considering one- and two-parameter POVMs.

\subsection{For one-parameter POVMs}
For one-parameter POVMs, we take the biasedness parameter $\alpha_{n,j} = 0$ i.e., $\xi_{n,j}=\sqrt{1-\eta^2_{n,j}}$ and $\gamma_{n,j}=(1+(n-1)\sqrt{1-\eta^2_{n,j}})/n$. For $n=2$, Bell inequality in  Eq.(\ref{localbound}) reduces to CHSH inequality. For a maximally entangled state  shared between Alice and first Bob, the lower bound of sharpness parameter $\eta_{2,j}$ of first Bob required for violating local bound of CHSH inequality is given by 
 $\eta_{2,1}  > \frac{1}{\sqrt{2}}$ and for second Bob
 \begin{eqnarray}
 \eta_{2,2}  > \frac{2\binom{1}{\lfloor\frac{1}{2}\rfloor}}{ \sqrt{2} (1+ \sqrt{1-\eta^2_{2,1}})}  \approx 0.83  ,
\end{eqnarray}
respectively. However, sharpness parameter $\eta_{2,3}$ of third Bob required for violating local bound of CHSH inequality is found to be $1.06$, which is not a legitimate value. The quantum value of CHSH expression for first and second Bob are $2.82$ and $2.41$ respectively and there is no violation of CHSH inequality for third Bob within valid range of $\eta_{2,3}$. This result is in accordance with Silva \emph{et al.} \cite{silva}. 

We then consider $n=3$ case for which the Bell inequality in Eq.(\ref{localbound}) becomes Gisin's elegant Bell inequality. Putting $n=3$ in Eq.(\ref{lvcond}) the lower bound of sharpness parameter of first and second Bob is obtained as
 $\eta_{3,1}  > \frac{6}{4\sqrt{3}} \approx 0.87$ and 
 $ \eta_{3,2} > 1.31$ respectively. Since $\eta_{3,2}$ is again outside the legitimate range of sharpness parameter, sharing of non-locality by second Bob through the violation of Gisin's elegant Bell inequality is not possible. Thus, for one-parameter POVMs, only one Bob can share non-locality for $n=3$. Following the similar procedure adopted above, we found that when $n\geq 3$ at most one Bob can share non-locality.  In Figure 1 points represents the minimum value of sharpness parameter of first Bob obtained for the violation of family of Bell's inequality given in Eq.(\ref{nbell1}) upto $n=100$.
\begin{figure}[ht]
\includegraphics[width=1\linewidth]{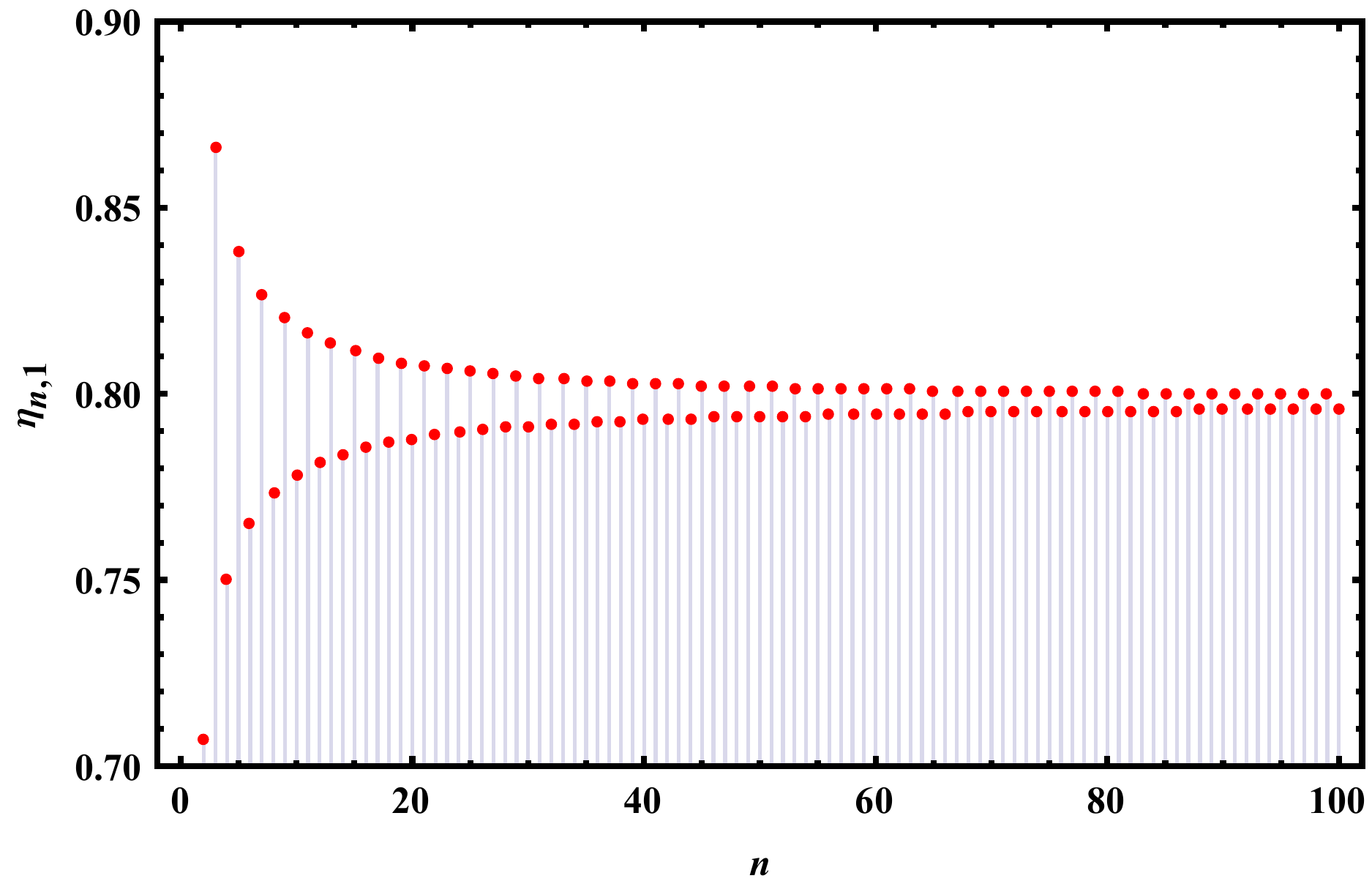}
\caption{(color online): Minimum value of sharpness parameter of first Bob for the violation family of Bell's inequality obtained upto $n=100$ using one-parameter POVMs}
\label{fig:01}
\end{figure}
It is seen from Figure 1 that the Bell's inequalities in Eq.(\ref{nbell1})  is violated for first Bob, and while $n$ increases the sharpness parameter of first Bob is approximately saturated to $0.82$. Points in Figure 2 represents the minimum value of sharpness parameter of second Bob required for the violations of the inequalities in Eq.(\ref{nbell1}) upto $n=100$.  
\begin{figure}[ht]
\includegraphics[width=1\linewidth]{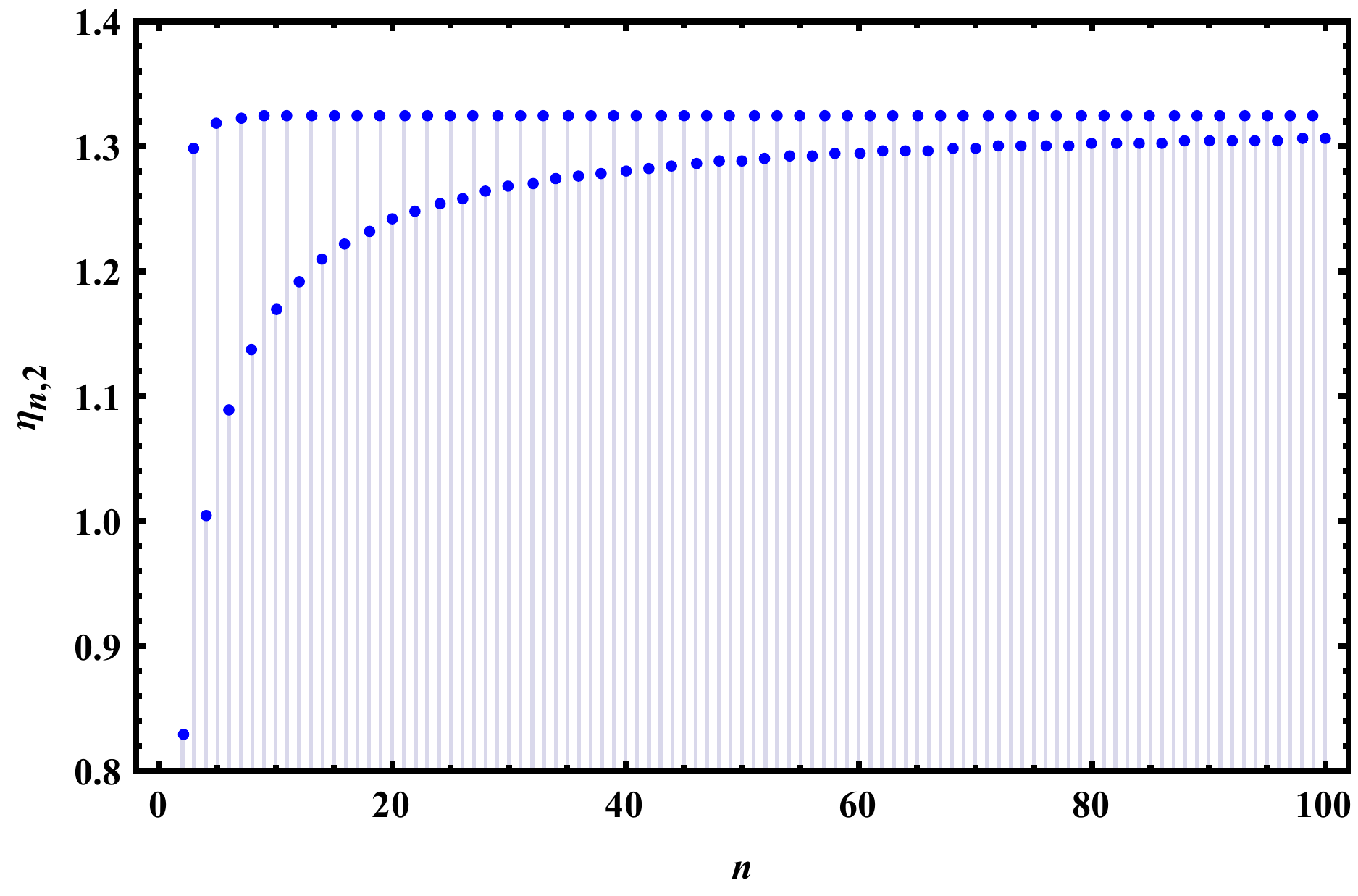}
\caption{(color online): Minimum value of sharpness parameter of second Bob obtained for the violation family of Bell's inequality upto $n=100$ using one-parameter POVMs.}
\label{fig:02}
\end{figure}
It is seen that except for $n=2$, the sharpness parameter of second Bob does not lie within valid range. Next, we demonstrate that the sharing of non-locality is also possible for second Bob if a particular two-parameter POVMs are taken for $n=2$, which was not studied earlier.

\subsection{For two-parameter POVMs}
Let us consider that each Bob performs sequential unsharp measurement described by two-parameter POVMs satisfying $|\alpha_{n,j}|+\eta_{n,j} \leq 1$. We can then have following two cases:\\

$\textit{Case(i)} : |\alpha_{n,j}| + \eta_{n,j} = 1$ \\

Substituting $|\alpha_{n,j}|  = 1 - \eta_{n,j}$, we have $\xi_{n,j}=\sqrt{1-\eta_{n,j}}$ and $\gamma_{n,j}=(1+(n-1)\sqrt{(1-\eta_{n,j}})/n$. The lower bound of unsharpness parameter of first and second Bob required for violating local bound of CHSH inequality are given by 
$\eta_{2,1}  >  \frac{1}{\sqrt{2}}$ and 
 \begin{eqnarray}
 \eta_{2,2}  >  \frac{2\binom{1}{\lfloor\frac{1}{2}\rfloor}}{ \sqrt{2} (1+ \sqrt{1-\eta_{2,1}})}\approx 0.92  ,
\end{eqnarray}
But third Bob requires  $ \eta_{2,3} \geq 1.42$, which is not a legitimate value. Thus, non-locality cannot be shared by third Bob. \\

$\textit{Case(ii)} : |\alpha_{n,j}| + \eta_{n,j} < 1$ \\

We take a particular case when $|\alpha_{n,j}| + \eta_{n,j} = 0.85$. The lower bound of unsharpness parameter of first and second Bob required for violating local bound of CHSH inequality is 
$ \eta_{2,1}  >  \frac{1}{\sqrt{2}}$ i.e., ($|\alpha_{n,j}| =0.14$) and for second Bob we take $|\alpha_{n,j}| =0.02$, we found

 \begin{eqnarray}
 \eta_{2,2} > \frac{2\binom{1}{\lfloor\frac{1}{2}\rfloor}}{2 \sqrt{2} \gamma_1 }  \approx 0.83,
\end{eqnarray}
But third Bob requires $\eta_{2,3}  > 1$ which is again not a legitimate value of sharpness parameter. The quantum value of $(\mathcal{B}_2)_Q$ for first and second Bobs are $2.82$ snd $2.39$ respectively. We have numerically checked that for two-parameter POVMs satisfying $\alpha_{2,j}| + \eta_{2,j} < 1$, there is no violation of CHSH inequality for third Bob.

 We have thus shown that for both one and two-parameter POVMs the non-locality can be shared for at most two Bobs through the violation of CHSH inequality. However, we have checked that for $n\geq 3$ at most one Bob can share non-locality using family of Bell's inequality given in Eq.(\ref{localbound}) for both forms of POVMs. We shall shortly demonstrate that the non-trivial preparation contextuality can be shared by arbitrary many Bobs for both forms of POVMs.

\section{Sharing non-trivial preparation contextuality for $k^{th}$ sequential Bob}
We shall now examine how many Bob can sequentially share non-trivial preparation contextuality when Bob's inputs are unbiased. The quantum violation of non-trivial preparation non-contextuality can be obtained for $k^{th}$ Bob, if $(\mathcal{B}^{k}_n)_{Q} >  (\mathcal{B}_n)_{pnc}$ is satisfied. Then, for $n$-bit case, from Eqs. (\ref{npnc}) and (\ref{quant}), the general condition on sharpness parameter 
 \begin{eqnarray}
\label{pn}
&& \eta_{n,k}  > \frac{1}{\sqrt{n}\prod\limits^{k-1}_{j=1} \gamma_{n,j}}
\end{eqnarray} 
 needs to be satisfied if non-trivial preparation contextuality is shared between Alice and $k^{th}$ Bob. Note that for first Bob ($k=1$) one needs $2^{n-1}\sqrt{n}\eta_{n,1}>2^{n-1} $ providing the condition on the sharpness parameter $\eta_{n,1} \geq 1/\sqrt{n} $. 

As already mentioned, for $n=2$, the Bell expression $(\mathcal{B}_2)$ is just the CHSH one. In this case no non-trivial constraint on Alice's observable exists,  and we have $(\mathcal{B}_2)_{local}=(\mathcal{B}_n)_{pnc}$. We then study the cases starting from $n\geq3$.
\subsection{For one-parameter POVMs}

 For $n=3$, if entangled state is shared between Alice and first Bob, then the lower bounds on sharpness parameter $\eta_{3,k}$ required for violating Gisin's elegant Bell inequality by first, second and third Bobs are given by $\eta_{3,1}  > \frac{1}{\sqrt{3}} = 0.57$, $\eta_{3,2} > 0.65 $  and $ \eta_{3,3} > 0.78$ respectively. However, sharpness parameter $\eta_{3,4}$ for fourth Bob is found to be $\eta_{3,4}>1.05$. Hence, there is no sharing of the non-trivial preparation non-contextuality for Alice and fourth Bob.

For $n=4$, the lower bound of unsharpness parameters required for first five Bobs are obtained as $\eta_{4,1} > 0.50$, $\eta_{4,2} > 0.56$, $\eta_{4,3} > 0.64$, $\eta_{4,4} > 0.77$ and $\eta_{4,5} > 1.05$. Again $\eta_{4,5}$ is  outside the valid range of sharpness parameter and hence sharing of non-trivial preparation contextuality by fifth Bob  for $n=4$ is not possible.

Let us now generalize the above findings for $n$-bit case. The task here to find the maximum number of Bobs who can share non-trivial preparation contextuality. Alternatively, for $k^{th}$ Bob what is the minimum value of $n$ (say, $n(k)$) is required for which the non-trivial preparation contextuality can be shared, where $k$ is arbitrary. We provide an analytical proof of the above question with legitimate approximation. 

From Eq. (\ref{pn}), we find that the critical value of sharpness parameter for $k^{th}$ Bob is $\eta_{n,k}  = 1/(\sqrt{n}\prod^{k-1}_{j=1} \gamma_{n,j})$ below which the non-contextual inequality given by Eq. (\ref{npnc}) will not be violated.  As shown earlier, the sharing of non-locality by  sequential Bobs is examined by considering the critical value of the previous Bobs. In fact, other analytical proofs \cite{silva,malar,das} for demonstrating the sharing non-locality have used this idea to find the legitimate unsharpness parameter for the sequential Bobs. 

If $k^{th}$ Bob shares the non-trivial preparation contextuality with Alice, the unsharpness parameter has to satisfy 
\begin{eqnarray}
\label{aa}
\eta_{n,k} \geq \frac{1}{\sqrt{n} \gamma_{n,k-1} \prod\limits^{k-2}_{j=1} \gamma_{n,j}}
\end{eqnarray}
Using the critical value of the unsharpness parameter of $(k-1)^{th}$ Bob so that $\eta_{n,k-1}  = 1/(\sqrt{n}\prod^{k-2}_{j=1} \gamma_{n,j})$, we can re-write the Eq. (\ref{aa}) as
\begin{eqnarray}
\label{aa11}
\eta_{n,k} \geq \frac{\eta_{n,k-1}} {\gamma_{n,k-1}} 
\end{eqnarray}

Now, we observe from Eq. (\ref{ga}) that $\gamma_{n,k-1}>\sqrt{1-\eta_{n,k-1}^2}$. Thus, $\frac{\eta_{n,k-1}}{\sqrt{1-\eta_{n,k-1}^2}}>\frac{\eta_{n,k-1}}{\gamma_{n,k-1}}$. Using this in Eq. (\ref{aa11}) one can demand the lower bound on $\eta_{n,k} $ requires
\begin{eqnarray}
\label{aa1}
\eta_{n,k} \geq  \frac{ \eta_{n,k-1}}{\sqrt{1-\eta^2_{n,k-1}}}
\end{eqnarray}

 By noting the critical value of sharpness parameter required for first Bob ($k=1$) $\eta_{n,1}= 1/\sqrt{n}$ and  by using Eq.(\ref{aa1}) we can estimate  the lower bound  $\eta_{n,2} \geq   1/\sqrt{n-1}$. Similarly, for third Bob ($k=3$), putting the critical values of  $ \eta_{n,1}$ and  $ \eta_{n,2}$ in Eq.(\ref{aa1}), we get $\eta_{n,3} \geq 1/\sqrt{n-2}$. Note that, Eq.(\ref{aa1}) is approximated and for small $n$ it overestimates the lower bound of the sharpness parameter. We can then generalize the argument for any arbitrary number of $k$. Thus, if ${k}^{th}$ Bob shares non-trivial preparation contextuality with Alice, then for $n$-bit case the following condition on sharpness parameter 
\begin{eqnarray}
\eta_{n,k} \geq \frac{1}{\sqrt{n-k+1}}
\end{eqnarray}
has to be satisfied. Alternatively, for a given $k$, one can find a $n\equiv n(k)$ for which preparation contextuality can be shared by $k^{th}$ Bob is given by
\begin{eqnarray}
n(k) \geq k-1 + \frac{1}{\eta^2_{n,k} }
\end{eqnarray}
If  the final Bob (${k}^{th}$) performs sharp measurement (i.e., $\eta_{n,k} = 1)$, we then have 
\begin{eqnarray}
\label{nf}
n(k) \geq k 
\end{eqnarray}
Thus, for any arbitrary number $(k)$ of Bob, there exists a $n(k)$ equal to or greater than $k$ for which the non-trivial preparation contextuality can be shared.

\subsection{For two-parameter POVMs}

We now proceed to examine whether preparation contextuality can be shared for an arbitrary sequences of Bob by using two-parameter POVMs. Intuitively, it can be argued that, in our case, the one-parameter POVMs performs better than two-parameter POVMs satisfying $|\alpha_{n,j}| + \eta_{n,j} \leq 1 $. If for a two-parameter POVMs, the relation between sharpness and biasedness parameters satisfy $|\alpha_{n,j}| + \eta_{n,j} = 1 $, then the value of  $|\alpha_{n,j}| $ is fixed by value of $\eta_{n,j}$. However, if  $|\alpha_{n,j}| + \eta_{n,j} < 1 $ one has the flexibility to choose the value of $|\alpha_{n,j}|$. For a given $n$ by taking the lower values of $|\alpha_{n,j}|$ , the number of Bobs sharing preparation contextuality can be made larger. Maximum number of Bob can share preparation contextuality when $|\alpha_{n,j}|$ approaches zero, which is one-parameter POVMs. We explicitly demonstrate this intuitive observation through numerical calculation by considering $n$ upto $100$.

Since $|\alpha_{n,j}| + \eta_{n,j} \leq 1$, let us consider the following two cases. \\

$\textit{Case(i)} : |\alpha_{n,j}| + \eta_{n,j} = 1$ \\

For $n=3$, the lower bound of sharpness parameter of required for first and second Bobs for sharing non-trivial preparation cotextuality are given by $\eta_{3,1}  > \frac{1}{\sqrt{3}} = 0.57$ and
\begin{eqnarray}
\eta_{3,2}  && >  \frac{3}{ \sqrt{3} (1+ 2 \sqrt{1-\eta_{3,1}})} \approx 0.75  \end{eqnarray}
However, sharpness parameter $\eta_{3,3} $ of third Bob is found to be $1.13$, which is not a legitimate value. Hence, if $|\alpha_{n,j}| + \eta_{n,j} = 1$ there is no violation of non-trivial preparation non-contextuality of Gisin's elegant Bell inequality for third Bob. 

Following the steps adopted for one-parameter POVMs, let us now find out  maximum how many number of Bob ($k$) can share non-trivial preparation contextuality in $n$-bit case for the two-parameters POVMs satisfying $|\alpha_{n,j}| + \eta_{n,j} = 1$. In that case, $\gamma_{n,j} = (1+(n-1)\sqrt{1-\eta_{n,j}})/n $. Now, using $\gamma_{n,k} \geq \sqrt{1-\eta_{n,k}}$ for large $n$ limit the condition for violating non-trivial preparation contextual bound by ${(k+1)}^{th}$ Bob is given by
\begin{eqnarray}
\label{aa2}
  \eta_{n,k+1} >  \frac{ \eta_{n,k}}{\sqrt{1-\eta_{n,k}}}
\end{eqnarray}
 First Bob ($k$) shares non-trivial preparation contextuality if $\eta_{n,1} \geq 1/\sqrt{n} $ is satisfied. The condition on the unsharpness parameter for sharing non-trivial preparation contextuality by $k^{th}$ is given by
\begin{eqnarray}
\label{kp1}
\nonumber
\eta_{n,k} > && \bigg[(\eta_{n,2})^{2^{k-1}}\bigg( (1-\eta_{n,k})(1-\eta_{n,k-1})^{2} (1-\eta_{n,k-2})^{2^2} \\  && (1-\eta_{n,k-3})^{2^3}......(1-\eta_{n,2})^{2^{k-2}}\bigg)^{-1}\bigg]^{-1/2}
\end{eqnarray}
From Eq.(\ref{kp1}) one finds that sharing of preparation contextuality depends on unsharpness parameter of all the previous Bobs.
\begin{figure}[ht]
\includegraphics[width=0.9\linewidth]{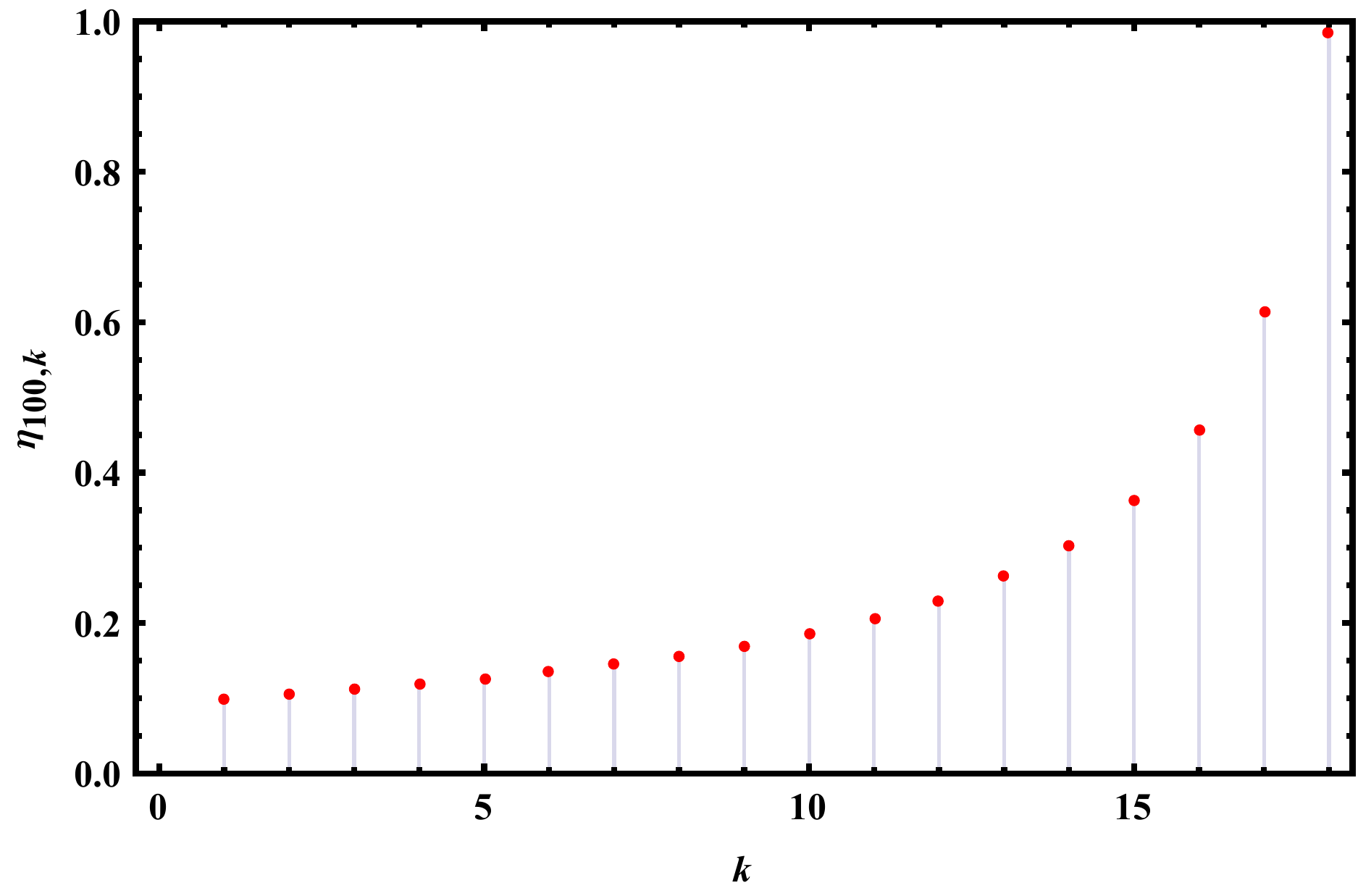}
\caption{(color online) :  Minimum value of sharppness parameter of $18$ Bobs violating family of Bell's inequality for $n=100$ given $\alpha_{n,j}+\eta_{n,j} = 1$.}
\label{fig:04}
\end{figure}
Figure 3 shows that for $n=100$, maximum $18$ Bobs can share non-trivial preparation contextuality. But sharpness parameter of $19^{th}$ Bob found to be outside the valid range of sharpness parameter.  However, the situation improves if we take $ |\alpha_{n,j}| + \eta_{n,j} < 1$. Note here that for the first Bob the quantity $(\mathcal{B}^{1}_n)_{Q}$ is independent of $|\alpha_{n,j}| $ and thereby depends only on $\eta_{n,j}$ and quantum violation of preparation contextuality obtained for $\eta_{n,1}>1/\sqrt{n}$. Thus, $|\alpha_{n,1}| $ can take value $0<\alpha_{n,1}<1-1/\sqrt{n}$. This actually fixes the range of allowed value of $|\alpha_{n,j}| $ when $j>2$.\\

$\textit{Case(ii)} : |\alpha_{n,j}| + \eta_{n,j} < 1$ \\

For $n=3$ the sharpness parameter of first Bob is $\eta_{3,1}  > \frac{1}{\sqrt{3}} \approx 0.57 $, that fixes the value of $|\alpha_{n,j}| < 0.43$. Further, we take $\alpha_{n,j} = 0.18$ with $\xi_{n,j} = (\sqrt{(1+\alpha_{n,j})^2-\eta_{n,j}^2}+\sqrt{(1-\alpha_{n,j})^2-\eta^2_{n,j}})/2$. In that case, second and third   Bobs share non-trivial preparation contextuality as
\begin{eqnarray}
\nonumber
 \eta_{3,2}  && >  \frac{3}{ \sqrt{3} (1+ 2 \xi_{3,1})} \approx 0.66   
\end{eqnarray}
and
 \begin{eqnarray}
\nonumber
&& \eta_{3,3}  \geq  \frac{9}{ \sqrt{3} (1+ 2 \xi_{3,1} )(1+ 2  \xi_{3,2})} \approx 0.80   
\end{eqnarray} 
But, for fourth Bob $\eta_{3,4} \geq 1.22 $ is required,   which does not lie in valid range. Hence for this particular two-parameter POVMs, the preparation contextuality can be shared by at least three Bobs sequentially, similar to the case of one-parameter POVMs.
 
\begin{figure}[ht]
\includegraphics[width=0.9\linewidth]{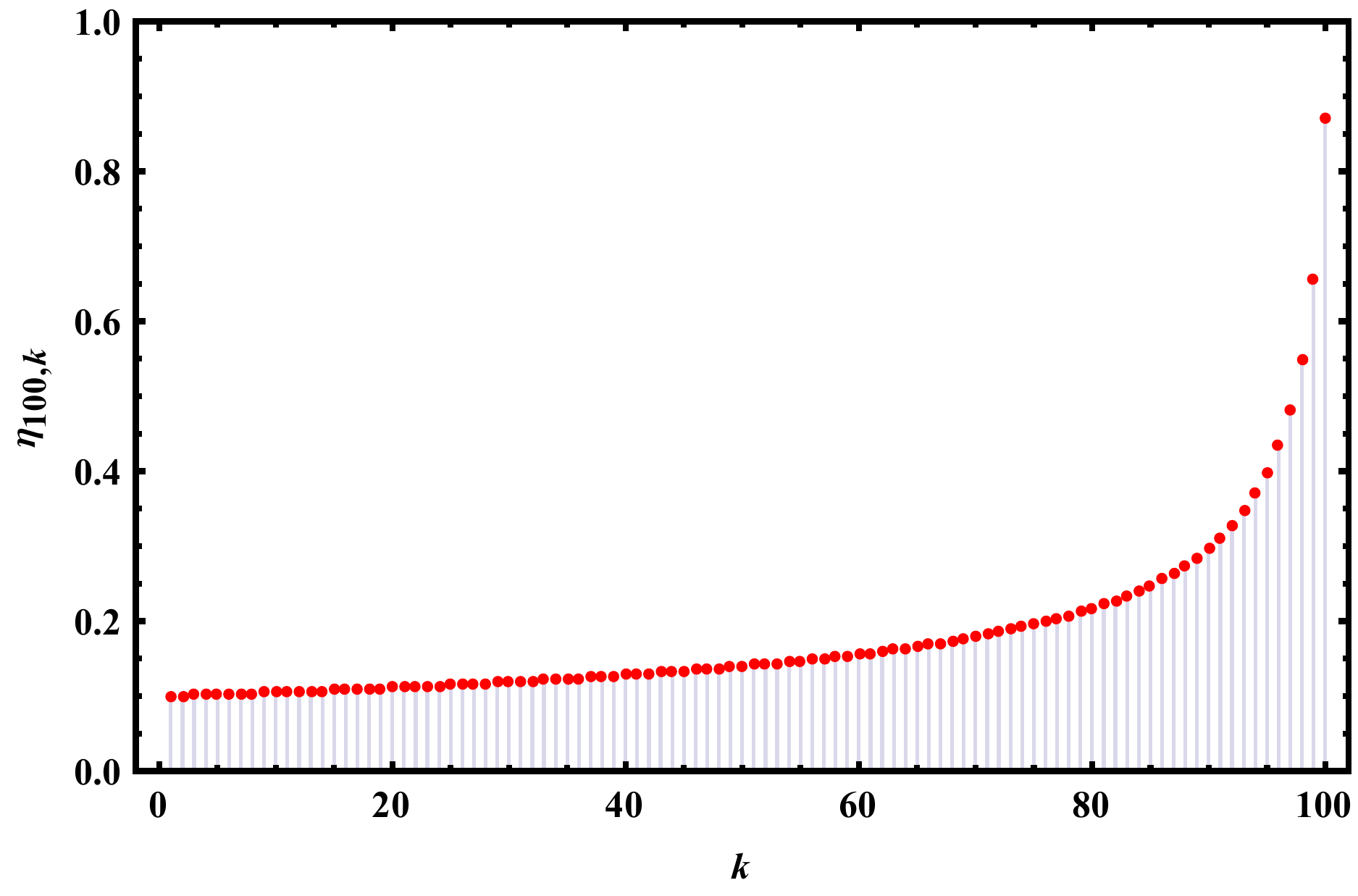}
\caption{(color online): Minimum value of sharpness parameter ($\eta_{100,k}$) of $100$ Bobs violating family of Bell's inequality for $n=100$, given $\alpha_{n,j}+\eta_{n,j} < 1$ with $\alpha_{100,k} = 0.08$ .}
\label{fig:04}
\end{figure}
 Further, we have numerically showed that for the case of $n=100$ by choosing $\alpha_{100,k}=0.08$, upto $100$ Bobs can sequentially share preparation contextuality.  In Figure 4 we have shown that similar to one-parameter POVMs for the case of two-parameter POVMs, $100$ Bobs can shares preparation contextuality if $n=100$. We  have numerically checked that the non-trivial preparation contextuality can be shared by any arbitrary number of Bob for a specific choices of two-parameter POVMs.

\section{Discussion regarding the biased choice of measurement settings by Bob }
We note here that throughout our paper we have considered unbiased measurement settings of Bobs. For two-qubit system, Silva \emph{et al.}\cite{silva} have shown that for biased choices of measurement settings by Bob, the non-locality can be sequentially shared by arbitrary number of Bobs through the  violation of CHSH inequality. For $n=2$ the Bell expression given by Eq.(\ref{nbell1}) is just the CHSH expression. As mentioned earlier,  each sequential Bob chooses the same two observables $B_{2,1}$ and $B_{2,2}$ that are used by first Bob. If the choices of measurement settings in CHSH scenario are biased, then while the first Bob chooses  $B_{2,1}$,  the second Bob may choose  the measurement of $B_{2,1}$ on the reduced state obtained from first Bob's measurement with probability $p_{2,2}$, and $B_{2,2}$ with probability $1-p_{2,2}$. Similarly,  when first Bob performs the measurement of $B_{2,2}$ the second Bob may choose $B_{2,2}$ with probability $p_{2,2}$ and $B_{2,1}$ with probability $1-p_{2,2}$. In biased choice of measurement settings, each Bob (starting from second Bob) sequentially measures same observable  with probability $p_{2,j}$ but different observable with probability  $1-p_{2,j}$ with $j\geq2$ where $j$ is the number of Bobs. 

Then, for the maximally entangled state given in Eq.(\ref{ent}), the quantum value of CHSH expression for $k^{th}$ Bob is obtain as
\begin{eqnarray}
\nonumber
(\mathcal{B}^{k}_2)_{Q} &=&  2\sqrt{2} \ \eta_{2,k} \prod^{k-1}_{j=1} \Big [p_{2,j}+(1-p_{2,j})\sqrt{1-\eta^2_{2,j}}\Big]
\end{eqnarray}
For unbiased case when all $p_{2,j}= 1/2$, the sharing of non-locality can be obtained for only for two Bobs. But with increasing value of $p_{2,j} > 1/2$ number of Bobs sharing non-locality increases and when $p_{2,j}$ is very close to one, very large number of Bobs can share non-locality with Alice. In one of the two extreme cases, when all $p_{2,j}=1$, we have $(\mathcal{B}^{k}_2)_{Q} =  2\sqrt{2} \ \eta_{2,k} $ implying that for any arbitrary $k$ if $\eta_{2,k}>1/\sqrt{2}$, one has the violation of CHSH inequality. In the other extreme case, when $p_{2,j}=0$, we have $(\mathcal{B}^{k}_2)_{Q} =  2\sqrt{2} \ \eta_{2,k} \prod\limits^{k-1}_{j=1} \sqrt{1-\eta^2_{2,j}}$. Since $\eta_{2,1}>1/\sqrt{2}$ is required to be satisfied for the first Bob, the sharing of non-locality cannot be demonstrated for more than one Bob. Intuitively,  when all $p_{2,j}=1$ with $j\geq2$, each Bob actually measures the same observable in sequence for every $j$. For example, second Bob measures $B_{2,1}$ only when the first Bob measures the same observable $B_{2,1}$. Then, the prior measurement will not disturb the state for the future measurements and only unsharpness parameter of the final measurement of Bob will appear in the CHSH expression, as seen above. 

The above argument can be generalized for any arbitrary $n$. For this, let  each Bob (starting from second Bob) sequentially measures same observable  with probability $p_{n,j}$, but each of the $(n-1)$  different observables with the average probability  $(1-p_{n,j})/(n-1)$. Due to such bias in choosing the measurement settings by Bob, the maximum quantum value of Bell expression given by Eq. (\ref{nbell1}) for $k^{th}$ Bob can be written as 
\begin{eqnarray}
\nonumber
(\mathcal{B}^{k}_n)_{Q} &=&  2^{n-1}\sqrt{n} \ \eta_{n,k} \prod^{k-1}_{j=1} \Big [p_{n,j}+(1-p_{n,j})\sqrt{1-\eta^2_{2,j}}\Big]
\end{eqnarray}
If the measurement settings are completely random, by putting $p_{n,j}=1/n$, the result of the unbiased settings given by Eq. (\ref{quant}) can easily be recovered. Following the argument provided for CHSH case, it is thus straightforward to understand that for biased choice of measurement settings of Bob, any arbitrary number of Bob can share non-trivial preparation contextuality for any given value of $n>3$. \\

\section{Summary }

We provided a detailed study of sharing of non-locality and non-trivial preparation contextuality through a family of Bell's expressions given by Eq. (\ref{nbell1}) whose local bound is considerably higher than non-trivial preparation non-contextual bound. As already mentioned that any proof of non-locality is a proof of trivial preparation contextuality \cite{pusey,barrett}. Note that, for $n=2$ case, the local and preparation contextual bound are the same. This is due to the fact that if Alice performs only two observables, no non-trivial relation can be established between those two observables \cite{ghorai18}. However, for $n\geq 3$ cases, a set of non-trivial relations given by Eq.(\ref{oba}) between Alice's observables can be found. Using such constraints the local bound can be considerably reduced which we termed as non-trivial preparation non-contextual bound. Thus, non-trivial preparation contextuality is a weaker form of correlation which may be reproduced by a local model.

 In order to find maximum how many number $(k)$ of Bobs can share non-locality and non-trivial preparation contextuality, we consider that Alice performs  projective measurements and $k-1$ number of Bobs sequentially perform unsharp measurement described by one- or  two-parameter POVMs. However, $k^{th}$ Bob measurement can be taken as sharp. We also considered that the choices of measurement settings by sequential Bobs are completely random. Since non-trivial preparation non-contextual bound of the aforementioned family of inequalities is lower than local bound then there is a possibility of sharing non-trivial preparation contextuality for more number of Bobs than that of non-locality.  We may then note that the violation of local bound warrants the violation of non-trivial preparation contextuality but converse may not be true in general. It is already known \cite{silva} that for the case of CHSH ($n=2$ in our case) inequality maximum two Bob can share non-locality. We showed that for two-parameter POVMs the inference remains same. We have also found that for $n\geq 3$, at most one Bob can share non-locality through the violation of the family of Bell's inequalities given by Eq. (\ref{localbound}).  

Further, through the quantum violation of non-trivial preparation non-contextual inequalities given by Eq.(\ref{npnc}), we have demonstrated that any arbitrary number of Bobs ($k$) can sequentially share non-trivial preparation contextuality for both one-parameter and for specific two-parameter POVMs. For this, the value of $n$ needs to be equal or greater than $k$.  We have also investigated the effect of biasedness in the choices of measurement settings by Bob and found that  any arbitrary number of Bobs can share non-trivial preparation contextuality for any given value of $n>3$. We note that, in \cite{shenoy} it was shown that quantum steering can be shared by arbitrary number of Bob. It is then tempting to find a connection between quantum steering and preparation contextuality. Such a result is demonstrated in a separate paper \cite{pan19}. 

Finally, we note here again that non-trivial preparation contextuality is a weaker form correlation than non-locality. Although non-trivial preparation contextuality can be shared by any arbitrary number of sequential Bobs in the case of unbiased input scenario, but the  sharing of non-locality by more than two Bob has not yet been demonstrated in the literature. It would then be interesting to study or to formulate suitable local realist inequalities to investigate whether  non-locality can also be shared by arbitrary number of Bobs in unbiased input scenario.  It is also worthwhile to explore the possibility of sharing non-locality for multi-outcome and multi-partite local realist inequalities. Studies along this line could
be an interesting avenue of research that will be carried out in
the future.

\textbf{Note added.-} During the completion of this work, we became aware of a similar work \cite{anwer19}.

\vskip -0.9cm
\section*{Acknowledgments}
 AKP acknowledges the support from the project DST/ICPS/QuEST/2018/Q-42.

\end{document}